*life*

*Article*

# Indexing Exoplanets with Physical Conditions Potentially Suitable for Rock-Dependent Extremophiles

**Madhu Kashyap Jagadeesh** [1,\*], **Sagarika Rao Valluri** [2], **Vani Kari** [3], **Katarzyna Kubska** [4] **and Łukasz Kaczmarek** [4]

[1] Department of Physics, Christ (Deemed to be university), Bengaluru, Karnataka 560029, India;
[2] Department of Electronics and Communication, RNSIT, Bengaluru, Karnataka 560098, India; sagarikarao1999@gmail.com
[3] Department of Biotechnology, Sri Jayachamarajendra College of Engineering, Mysuru, Karnataka, India 570006; vanikari547@gmail.com
[4] Department of Animal Taxonomy and Ecology, Faculty of Biology, Adam Mickiewicz University, Poznań, Uniwersytetu Poznańskiego 6, Poznań 61-614, Poland; katarzyna.qbska@gmail.com (K.K.); kaczmar@amu.edu.pl (L.K.)
\* Correspondence: kas7890.astro@gmail.com

**Abstract:** The search for different life forms elsewhere in the universe is a fascinating area of research in astrophysics and astrobiology. Currently, according to the NASA Exoplanet Archive database, 3876 exoplanets have been discovered. The Earth Similarity Index (ESI) is defined as the geometric mean of radius, density, escape velocity, and surface temperature and ranges from 0 (dissimilar to Earth) to 1 (similar to Earth). The ESI was created to index exoplanets on the basis of their similarity to Earth. In this paper, we examined rocky exoplanets whose physical conditions are potentially suitable for the survival of rock-dependent extremophiles, such as the cyanobacteria *Chroococcidiopsis* and the lichen *Acarospora*. The Rock Similarity Index (RSI) is first introduced and then applied to 1659 rocky exoplanets. The RSI represents a measure for Earth-like planets on which physical conditions are potentially suitable for rocky extremophiles that can survive in Earth-like extreme habitats (i.e., hot deserts and cold, frozen lands).

## 1. Introduction

In recent years, extraterrestrial research has become the 'holy grail' of astrobiology. Space missions like CoRoT (Convection, Rotation and planetary Transits) and Kepler have provided a huge amount of data from exoplanetary observations which are catalogued in the Planetary Habitability Laboratory, (PHL-EC, University of Puerto Rico (UPR), Arecibo, 2017, http://phl.upr.edu/projects/habitable-exoplanets-catalogue/data/database) [1]. The PHL-EC data (as of 2018) for different planetary objects, such as radius, density, escape velocity, and surface temperature, have been used to create a metric index called the Earth Similarity Index (ESI) that ranges from 0 (dissimilar to Earth) to 1 (identical to Earth) [2]. The ESI allows Earth-like and potentially habitable planets (PHPs) to be identified on the basis of the observed physical parameters of extra-solar objects.

Exoplanets can be divided into rocky planets of different sizes and gas giants. The masses of rocky planets range from 0.1 to 10 Earth masses, while the radii range from 0.5 to 2 Earth radii [3]. Recently, Kashyap et al. [4] introduced a new technique to estimate the surface temperature of different exoplanets and formulated the Mars Similarity Index (MSI) for the search for extremophilic life forms which are capable of survival in Mars-like conditions.

In 2018, Kashyap et al. [5] introduced two additional indexes: the Active Tardigrade Index (ATI) and the Cryptobiotic Tardigrade Index (CTI). Both the ATI and CTI were designed to catalogue exoplanets according to the potential survivability of extremophilic invertebrates (e.g., Tardigrada (water bears)) on their surfaces. The ATI and CTI are defined as the geometric mean of radius, density, escape velocity, surface temperature, surface pressure, and revolution, in a range from 0 to 1. This paper focuses on rocky exoplanets with Earth-like conditions and surface temperatures varying within a range potentially suitable for growth and reproduction of extremophilic microorganisms.

Extremophiles are organisms which are able to survive extreme physical or geochemical conditions that are lethal, or at least harmful, to most organisms on Earth [6]. These organisms can be found in all kingdoms of life, but most of them belong to Bacteria and Archaea. In addition, such organisms can also be found among animals, fungi, and plants. The organisms considered to be the most tolerant include fungi, lichens, algae, tardigrades, rotifers, nematodes, and some insects and crustaceans [7–14]. This paper focuses on two extremophiles growing on rocks: the cyanobacteria *Chroococcidiopsis* and the lichen *Acarospora* [15–16].

*Chroococcidiopsis* is a photosynthetic primitive cyanobacteria growing on and below rocks and characterized by a high potential to colonization and recolonization of extreme habitats [17]. *Chroococcidiopsis* is known for its tolerance of harsh conditions, including high and low temperatures, ionising radiation, and high salinity [18]. Verseux et al. [19] proposed that *Chroococcidiopsis* is an organism capable of living on Mars and potentially capable of terraforming the red planet. Additionally, *Chroococcidiopsis* was used in tests involving low Earth orbit, impact events, planetary ejection, atmospheric re-entry, and simulated Martian conditions [20–23].

*Acarospora* species are crustose lichens inhabiting xerothermic habitats that grow on dry rocks [24] and tolerate harsh conditions such as low and high temperatures, high radiation, or lack of water [15,25]. Research has shown that two *Acarospora* species are capable of survival in a simulated Martian environment [26].

This paper introduces the Rock Similarity Index (RSI) and calculates RSI for 1659 rocky-iron exoplanets. The RSI is similar to the ATI and CTI (as calculated in [5]), yet differs in that the surface temperature parameter is modified to reflect the potential survivability of rock-dependent extremophiles.

Weight exponent calculation of Mars where the threshold value is:

$$V = \left[1 - \left|\frac{x - x_0}{x + x_0}\right|\right]^{w_x}$$

where $w_x$ is the weight exponent required, $X_0$ is defined as the reference value, and $X_a < X_0 < X_b$ (a and b are the upper and lower limits)

$$w_a = \frac{lnV}{ln\left[1 - \left|\frac{x_a - x_0}{x_a + x_0}\right|\right]} \quad (x_a = 258 \text{ K}, x_0 = 288 \text{ K}, V = 0.80)$$

$$w_a = \frac{ln 0.80}{ln\left[1 - \left|\frac{258 - 288}{258 + 288}\right|\right]}$$

$$w_a = 3.9485$$

$$w_b = \frac{lnV}{ln\left[1-\left|\frac{x_b-x_0}{x_b+x_0}\right|\right]} \quad (x_b = 395\ K,\ x_0 = 288\ K,\ V = 0.80)$$

$$w_b = \frac{ln0.80}{ln\left[1-\left|\frac{395-288}{395+288}\right|\right]} \quad w_b = 1.3096$$

$$w_x = \sqrt{w_a x w_b}$$

$$w_x = \sqrt{3.9485 x 1.3096}$$

$$w_x = 2.26$$

## 2. Results

The RSI is designed to index Earth-like planets with physical conditions which, though harsh, are at least potentially suitable for rock-dependent extremophiles such as *Chroococcidiopsis* and *Acarospora*. According to Mckay [27], generally speaking, the temperature range in which extremophilic microorganisms are able to reproduce and grow is between 258 K and 395 K. With regard to the calculation of the RSI, the corresponding weight exponent for surface temperature was calculated to be 2.26. We calculated the RSI average weight exponents for rocky exoplanets, as shown in Table 1.

**Table 1.** Parameters used to calculate the weight exponents for the Rock Similarity Index (RSI) scale.

| Planetary Property | Reference Values for RSI | Weight Exponents for RSI |
|---|---|---|
| Mean radius | 1 EU | 0.57 |
| Bulk density | 1 EU | 1.07 |
| Escape velocity | 1 EU | 0.70 |
| Surface temperature | 288 K | 2.26 |
| Surface pressure | 1 EU | 0.022 |
| Revolution | 1 Earth year | 0.7 |

The weight exponents for the upper and lower limits appeared similar to the tardigrade indexes of Kashyap et al. [5], with the exception of surface temperature. In order to calculate the surface temperature of the studied exoplanets, the albedo 0.3 (similar to that on Earth) was applied as a proxy (e.g., as seen in Table 2, for Proxima Cen b the effective temperature was 229.3 K, and the surface temperature was 263.9 K). In order to calculate the weight exponent, the following ranges were used for the upper and lower limits of each parameter: mean radius = 0.5–1.9 EU; bulk density = 0.7–1.5 EU; escape velocity = 0.4–1.4 EU; surface temperature T = 258–395 K; and revolution = 0.61–1.88 EU. The weight exponents were calculated by applying these limits in the weight exponent equation previously proposed [5].

The RSI for rock-dependent extremophiles is defined as the geometrical mean of radius, density, escape velocity, and surface temperature of exoplanets, in a range from 0 to 1, where 0 indicates non-survival, and 1 represents survival.

Mathematically,

$$RSI = (RSI_R \times RSI_\rho \times RSI_{v_e} \times RSI_{T_s} \times RSI_{rev} \times RSI_p)^{\frac{1}{6}}$$

where $RSI_R$, $RSI_\rho$, $RSI_{Ts}$, $RSI_{Ve}$, $RSI_{rev}$, and $RSI_p$ represent the RSI values of radius, density, surface temperature, escape velocity, revolution (Earth years), and pressure, respectively. The RSI of each physical parameter is defined similarly to the ESI and is given by:

$$RSI_x = \left[1-\left|\frac{x-x_0}{x+x_0}\right|\right]^{w_x}$$

where x represents a physical parameter of the exoplanet (radius *R*, bulk density *ρ*, escape velocity $V_e$, surface temperature $T_s$, pressure *p*, or revolution *rev*), $x_0$ denotes the reference value for Earth, and $w_x$ is the weight exponent, as seen in Table 1. Most parameters are expressed in EU (Earth units), while the surface temperature is given in Kelvin (K).

The global RSI is divided into interior ($RSI_I$) and surface ($RSI_S$), which are expressed as:

$$RSI_I = (RSI_R \times RSI_\rho)^{\frac{1}{2}}$$

$$RSI_S = (RSI_{v_e} \times RSI_{T_s} \times RSI_{rev} \times RSI_p)^{\frac{1}{4}}$$

Therefore, the global RSI is defined as

$$RSI = (RSI_I \times RSI_S)^{\frac{1}{2}}$$

The RSI values are computed from Equations 2–5 using data from [4] for the radius, density, escape velocity, surface temperature, revolution, and pressure, together with the surface temperature weight exponent value of 2.26. A representative sample is shown in Table 2; the entire table is catalogued and made available online (see [28]).

**Table 2.** RSI analysis for Mars and various sample exoplanets compared to Earth where *R* = radius, *ρ* = density, *T* = surface temperature, *Ve* = escape velocity, *P* = pressure, Rev = revolution, $RSI_I$ = Interior Rock Similarity Index, $RSI_S$ = Surface Rock Similarity Index, RSI = Global Rock Similarity Index.

| Planet | R(EU) | ρ(EU) | T (K) | Ve(EU) | P(EU) | Rev(days) | $RSI_I$ | $RSI_S$ | RSI |
|---|---|---|---|---|---|---|---|---|---|
| Earth | 1.00 | 1.00 | 288 | 1.00 | 1.00 | 1.00 | 1.00 | 1.00 | 1.00 |
| Mars | 0.532 | 0.713 | 218 | 0.45 | 0.99 | 0.97 | 0.81 | 0.83 | 0.82 |
| Proxima Cen.-b | 1.12 | 0.9 | 263.9 | 0.97 | 0.99 | 0.14 | 0.95 | 0.59 | 0.75 |
| GJ 667Cc | 1.4 | 0.99 | 286.4 | 1.39 | 2.7 | 0.24 | 0.92 | 0.66 | 0.78 |
| Kepler-296e | 1.48 | 1.03 | 306.6 | 1.07 | 1.1 | 0.27 | 0.93 | 0.68 | 0.79 |

A graphical representation of rocky planets characterized according to the RSI is presented in Figure 1.

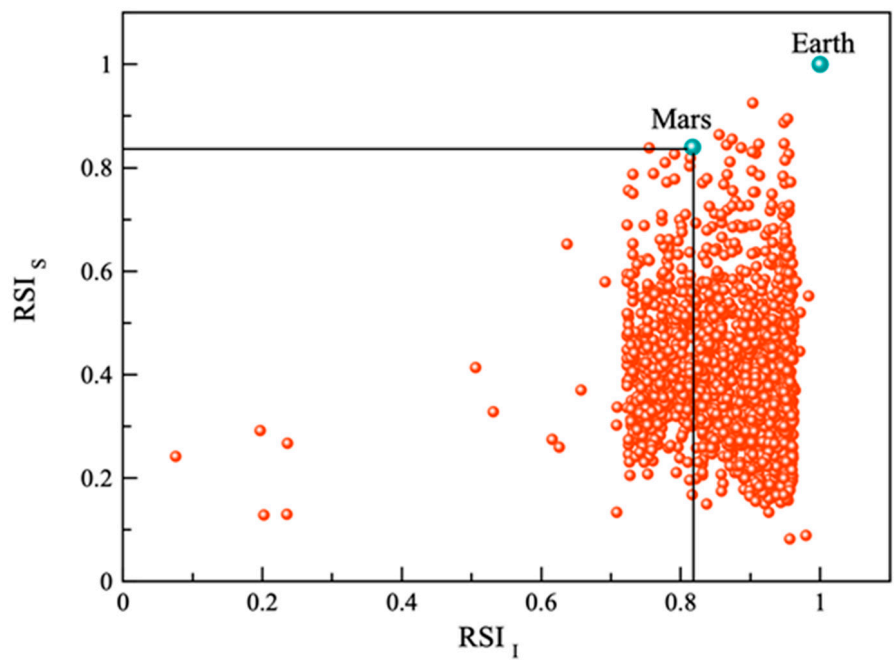

**Figure 1.** Scatter plot of surface RSI versus interior RSI. The solid line marked for Mars is the threshold value for potentially habitable planets.

The threshold (a limit for potential microorganisms survival) for rocky exoplanets that are considered to be potentially habitable by extremophiles such as *Chroococcidiopsis* and *Acarospora* is defined by considering Mars (on which this forms of life are able to survive [20,26]) that has an RSI of ~0.82 (for details see also calculations above).

## 3. Discussion and Conclusions

The search for extraterrestrial life forms has given rise to numerous space missions that have enabled researchers to collect data, test different species of extremophiles (e.g., black fungi, cyanobacteria, bryophytes, invertebrates) in space conditions, analyse their physiology [29] in extreme conditions, and finally find potentially habitable exoplanets for Earth-like organisms. Space missions which previously studied extremophiles include EXPOSE-E, EXPOSE-R2, BIOMEX, and CoRoT [30].

Up to now, Earth is the only known rocky planet which both has a developed biosphere and is shielded by a magnetic field that protects it against harmful cosmic radiation [31]. In this analysis, we focused on rocky exoplanets which have physical conditions similar to those of Earth or Mars. We chose two microorganisms, *Chroococcidiopsis* and *Acarospora*, that are able to survive, grow, and reproduce in very harsh conditions and in the absence of a planet's magnetic field. *Chroococcidiopsis* was previously selected for colonizing tests on Mars (Russian Expose Mission) because it can grow on rocks, produces oxygen, and tolerates high energy cosmic radiation [32]. Similarly, *Acarospora* was tested by the EXPOSE-E mission for one and a half years and managed to survive in Mars-like conditions [24].

According to Kashyap et al. [4], Mars, with an ESI value of 0.73, was defined as the limit for planets which could have physical conditions suitable for complex life forms. Based on this criterion, approximately 44 planets have been identified as PHPs. Considering the RSI for 1659 rocky exoplanets with a threshold of 0.82, 21 exoplanets have been found to be PHPs, where physical conditions are suitable for extremophiles such as *Chroococcidiopsis* and *Acarospora*.

A very important factor in our analysis is the calculation of the weight exponent for surface temperature. The weight exponents used for each physical factor allow an accurate calculation of the ESI and RSI, so it is crucial to have the correct weight exponent. For the calculation of the RSI, a temperature limit range from 273K to 373K was used [27], and the corresponding weight exponent for surface temperature was calculated to be 2.26. This value corresponds to the conditions which are potentially suitable for rock-dwelling extremophiles to survive. Subsequent space missions, such as the James Webb Space Telescope, will provide deeper insights into potentially habitable planets and their environments. Once the data from these missions have been combined with detailed knowledge on environmental conditions where extremophiles are potentially able to survive, it will be possible to identify potential physical and chemical parameters which should be present on exoplanets or exomoons to be suitable for Earth-like organisms. The RSI proposed by us is a tool which indexes planets that have physical conditions potentially suitable for certain Earth microorganisms. While it is obvious that our index does not provide definitive answers, it does enable us to identify the best candidate exoplanets or exomoons to be chosen for both further research and searches for extraterrestrial life signatures.

**Author Contributions:** conceptualization, M.K.J., V.K., and Ł.K.; methodology, M.K.J.; software, M.K.J. and S.R.V.; validation, M.K.J and Ł.K.; formal analysis, M.K.J.; investigation, M.K.J. and S.R.V.; resources, M.K.J.; data curation, M.K.J., S.R.V., and Ł.K.; writing—original draft preparation, M.K.J., S.R.V., V.K., K.K., and Ł.K.; writing—review and editing, M.K.J., S.R.V., V.K., K.K., and Ł.K.; visualization, M.K.J.; supervision, M.K.J and Ł.K. All authors have read and agreed to the published version of the manuscript.

**Funding:** This research received no external funding.

**Acknowledgments:** Special thanks to Vani, a former M. Tech Biotechnology project intern from the National Chemical Laboratory (NCL) Pune, for brief discussions of extremophiles. This research utilized the PHL-EC at UPR Arecibo (2017) (http://phl.upr.edu/projects/habitable-exoplanets-catalogue/data/database), the NASA Exoplanet Archive, and the NASA Astrophysics Data System Abstract Service.

**Conflicts of Interest:** The authors declare no conflict of interest. The parties who funded this study had no role in its design, in the collection, analyses, or interpretation of the data, or in the writing of the manuscript and the decision to publish the results.

**References**

1. Planetary Habitability Laboratory, PHL-EC@UPR Arecibo, 2017. Available online: http://phl.upr.edu/projects/habitable-exoplanets-catalog/data/database (accessed on June, 2018).
2. Schulze-Makuch, D.; Mndez, A.; Fairén, A.G.; von Paris, P.; Turse, C.; Boyer, G.; Davila, A.F.; António, M.R.; Catling, D.; Irwin, L.N. A two-tiered approach to assessing the habitability of exoplanets. *Astrobiology* **2011**, *11*, 1041–1052. doi:10.1089/ast.2010.0592.
3. Safonova, M.; Murthy, J.; Shchekinov, Y.M. Age aspects of habitability. *Int. J. Astrobiology* **2016**, *15*, 93–105. doi:10.1017/S1473550415000208.
4. Kashyap, J.M.; Gudennavar, S.B.; Doshi, U.; Safonova, M. Indexing of exoplanets in search for potential habitability: Application to Mars-like worlds. *Astrophys. Space Sci.* **2017**, *362*, 146. doi:10.1007/s10509-017-3131-y.
5. Kashyap, J.M.; Roszkowska, M.; Kaczmarek, Ł. Tardigrade indexing approach on exoplanets. *Life Sci. Space Res.* **2018**, *19*, 13–16. doi:10.1016/j.lssr.2018.08.001.
6. Rampelotto, P.H. Resistance of microorganisms to extreme environmental conditions and its contribution to Astrobiology. *Sustainability* **2010**, *2*, 1602–1623. doi:10.3390/su2061602.
7. McSorley, R. Adaptations of nematodes to environmental extremes. *Fla. Entomol.* **2003**, *86*, 138–142. doi:10.1653/0015-4040(2003)086[0138:AONTEE]2.0.CO;2.
8. Alpert, P. The limits and frontiers of desiccation-tolerant life. *Integr. Comp. Biol.* **2005**, *45*, 685–695. doi:10.1093/icb/45.5.685.
9. Islam, M.R.; Dirk Schulze-Makuch, D. Adaptations to environmental extremes by multicellular organisms. *Int. J. Astrobiology* **2007**, *6*, 199–215. doi:10.1017/S1473550407003783.
10. Guidetti, R.; Rizzo, A.M.; Altiero, T.; Rebecchi, L. What can we learn from the toughest animals of the Earth? Water bears (tardigrades) as multicellular model organisms in order to perform scientific preparations for lunar exploration. *Planet. Space Sci.* **2012**, *74*, 97–102. doi:10.1016/j.pss.2012.05.021.
11. Seckbach, J. Survey of algae in extreme environments. In *The Algae World. Cellular Origin, Life in Extreme Habitats and Astrobiology*; Sahoo, D., Seckbach, J., Eds.; Springer: Dordrecht, The Netherlands, 2015; Volume 26, pp. 307–315.
12. Armstrong, R.A. Adaptation of lichens to extreme conditions. In *Plant Adaptation Strategies in Changing Environment*; Shukla, V., Kumar, S., Kumar, N., Eds.; Springer: Singapore, 2017; pp. 1–27.
13. Onofri, S.; de la Torre, R.; de Vera, J.P.; Ott, S.; Zucconi, L.; Selbmann, L.; Scalzi, G.; Venkateswaran, K.J.; Rabbow, E.; Sánchez Iñigo, F.J.; et al. Survival of rock-colonizing organisms after 1.5 years in outer space. *Astrobiology* **2012**, *12*, 508–516. doi:10.1089/ast.2011.0736.
14. Selbmann, L.; Pacelli, C.; Zucconi, L.; Dadachova, E.; Moeller, R.; de Vera, J.P.; Onofri, S. Resistance of an Antarctic cryptoendolithic black fungus to radiation gives new insights of astrobiological relevance. *Fungal Biol.* **2018**, *122*, 546–554. doi:10.1016/j.funbio.2017.10.012.
15. Nash, T.H.; Ryan, B.D.; Gries, C.; Bugartz, F. Acarospora. In *Lichen Flora of the Greater Sonoran Desert Region*; Arizona State University: Tempe, AZ, USA, 2001.
16. Cumbers, J.; Rothschild, L.J. Salt tolerance and polyphyly in the cyanobacterium *Chroococcidiopsis* (Pleurocapsales). *J. Phycol.* **2014**, *50*, 472–482. doi:10.1111/jpy.12169.
17. Kolda, A.; Petrić, I.; Mucko, M.; Gottstein, S.; Žutinić, P.; Goreta, G.; Ternjej, I.; Rubinić, J.; Radišić, M.; Udovič, M.G. How environment selects: Resilience and survival of microbial mat community within intermittent karst spring Krčić (Croatia). *Ecohydrology* **2018**, *12*, e2063. doi:10.1002/eco.2063.
18. Wierzchos, J.; DiRuggiero, J.; Vítek, P.; Artieda, O.; Souza-Egipsy, V.; Škaloud, P.; Michel Tisza, M.; Davila, A.F.; Vílchez, C.; Garbayo, I.; et al. Adaptation strategies of endolithic chlorophototrophs to survive the hyperarid and extreme solar radiation environment of the Atacama Desert. *Front. Microbiol.* **2015**, *6*, 934. doi:10.3389/fmicb.2015.00934.
19. Verseux, C.; Baqué, M.; Lehto, K.; de Vera, J.-P.P.; Rothschild, L.J.; Billi, D. Sustainable life support on Mars—The potential roles of cyanobacteria. *Int. J. Astrobiology* **2015**, *15*, 65–92. doi:10.1017/S147355041500021X.

20. Cockell, C.S.; Schuerger, A.C.; Billi, D.; Friedmann, E.I.; Panitz, C. Effects of a simulated Martian UV flux on the cyanobacterium, *Chroococcidiopsis* sp. 029. *Astrobiology* **2005**, *5*, 127–140.
21. Cockell, C.S.; Rettberg, P.; Rabbow, E.; Olson-Francis, K. Exposure of phototrophs to 548 days in low Earth orbit: Microbial selection pressures in outer space and on early Earth. *ISME J.* **2011**, *5*, 1671–1682. doi:10.1038/ismej.2011.46.
22. Billi, D. Damage escape and repair in dried *Chroococcidiopsis* spp. from hot and cold deserts exposed to simulated space and Martian conditions. *Astrobiology* **2011**, *11*, 65–73. doi:10.1089/ast.2009.0430.
23. Baqué, M.; de Vera, J.; Rettberg, P.; Billi, D. The BOSS and BIOMEX space experiments on the EXPOSE-R2 mission: Endurance of the desert cyanobacterium *Chroococcidiopsis* under simulated space vacuum, Martian atmosphere, UVC radiation and temperature extremes. *Acta Astronaut.* **2013**, *91*, 180–186. doi:10.1016/j.actaastro.2013.05.015.
24. Favero-Longo, S.E.; Accattino, E.; Matteucci, E.; Borghi, A.; Piervittori, R. Weakening of gneiss surfaces colonized by endolithic lichens in the temperate climate area of northwest Italy, 2015. *Earth Surf. Proc. Land.* **2015**, *40*, 2000–2012. doi:10.1002/esp.3774.
25. Azua-Bustos, A.; Urrejola, C.; Vicuña, R. Life at the dry edge: Microorganisms of the Atacama Desert. *FEBS Lett.* **2012**, *586*, 2939–2945. doi:10.1016/j.febslet.2012.07.025.
26. Scalzi, G.; Selbmann, L.; Zucconi, L.; Rabbow, E.; Horneck, G.; Albertano, P.; Onofri, S. LIFE experiment: Isolation of cryptoendolithic organisms from Antarctic colonized sandstone exposed to space and simulated Mars conditions on the international space station. *Orig. Life Evol. Biosph.* **2012**, *42*, 253–262. doi:10.1007/s11084-012-9282-5.
27. McKay, C.P. Requirements and limits for life in the context of exoplanets. *PNAS* **2014**, *111*, 12628. doi:10.1073/pnas.1304212111.
28. Kashyap, J.M. "RSI", Mendeley Data; 2019. Available online: http://dx.doi.org/10.17632/xp4wy9px6n.1 (accessed on 26 January 2020).
29. de Vera, J.P.; Alawi, M.; Backhaus, T.; Baqué, M.; Billi, D.; Böttger, U.; Berger, T.; Bohmeier, M.; Cockell, C.; Demets, R.; et al. Limits of life and the habitability of Mars: The ESA space experiment BIOMEX on the ISS. *Astrobiology* **2019**, *19*, 145–157. doi:10.1089/ast.2018.1897.
30. Deleuil, M.; Fridlund, M. CoRoT: The first space-based transit survey to explore the close-in planet population. In *Handbook of Exoplanets*; Deeg, H., Belmonte, J., Eds.; Springer: Berlin/Heidelberg, Germany, 2018; pp. 1–22.
31. Evans, F. Electric and magnetic effects of cosmic rays. *Phys. Rev.* **1942**, *61*, 680. doi:10.1103/PhysRev.61.680.
32. Billi, D.; Friedmann, H.; Caiola, O. Ionizing-radiation resistance in the desiccation-tolerant cyanobacterium *Chroococcidiopsis*. *Appl. Environ. Microbiol.* **2000**, *66*, 1489–1492.